\documentclass[prl,twocolumn,showpacs,superscriptaddress]{revtex4}
\usepackage{epsf,graphicx,amssymb,mathptm,dcolumn}
\begin{document}

\title{Kinetic Roughening of Ion-Sputtered Pd(001) Surface: 
Beyond the Kuramoto-Sivashinsky Model}

\author{T.~C.~Kim}
\affiliation{Department of Materials Science and Engineering, 
Kwangju Institute of Science and Technology, Gwangju, 500-712, Korea}

\author{C.-M.~Ghim} 
\email[Electronic address: ]{cmghim@phya.snu.ac.kr}
\affiliation{School of Physics and Center for Theoretical Physics, 
Seoul National University, Seoul 151-747, Korea}

\author{H.~J.~Kim} 
\
\affiliation{Department of Materials Science and Engineering, 
Kwangju Institute of Science and Technology, Gwangju, 500-712, Korea}

\author{D.~H.~Kim}
\affiliation{Department of Materials Science and Engineering, 
Kwangju Institute of Science and Technology, Gwangju, 500-712, Korea}

\author{D.~Y.~Noh}
\affiliation{Department of Materials Science and Engineering, 
Kwangju Institute of Science and Technology, Gwangju, 500-712, Korea}

\author{N.~D.~Kim }
\affiliation{Department of Physics and Basic Science Research Institute, 
Pohang University of Science and Technology, Pohang 790-784, Korea}

\author{J.~W.~Chung }
\affiliation{Department of Physics and Basic Science Research Institute, 
Pohang University of Science and Technology, Pohang 790-784, Korea}

\author{J.~S.~Yang}
\affiliation{School of Physics and Research Center for Oxide Electronics, 
Seoul National University, Seoul 151-747, Korea}

\author{Y.~J.~Chang}
\affiliation{School of Physics and Research Center for Oxide Electronics, 
Seoul National University, Seoul 151-747, Korea}

\author{T.~W.~Noh}
\affiliation{School of Physics and Research Center for Oxide Electronics, 
Seoul National University, Seoul 151-747, Korea}

\author{B.~Kahng}
\affiliation{School of Physics and Center for Theoretical Physics, 
Seoul National University, Seoul 151-747, Korea}

\author{J.-S.~Kim}
\email[Electronic address: ]{jskim@sookmyung.ac.kr}
\affiliation{Department of Physics, Sook-Myung Women's University, 
Seoul 140-742, Korea}

\date{\today}

\begin{abstract}
We investigate the kinetic roughening of Ar$^+$ ion-sputtered 
Pd(001) surface both experimentally and theoretically. 
\textit{In situ} real-time x-ray reflectivity and \textit{in situ} 
scanning tunneling microscopy show that nanoscale adatom islands form 
and grow with increasing sputter time $t$. 
Surface roughness, $W(t)$, and lateral correlation length, 
$\xi(t)$, follows the scaling laws, $W(t)\sim t^{\beta}$ and 
$\xi(t)\sim t^{1/z}$ with the exponents $\beta\simeq 0.20$ and 
$1/z\simeq 0.20$, for ion beam energy $\epsilon=0.5$ keV,
which is inconsistent with the prediction of the Kuramoto-Sivashinsky 
(KS) model. 
We thereby extend the KS model by applying the Sigmund theory of sputter 
erosion to the higher order, ${\cal O}(\nabla^4, h^2)$, where $h$ is 
surface height, and derive a new term of the form $\nabla^2(\nabla h)^2$
which plays an indispensable role in describing the observed 
morphological evolution of the sputtered surface. 
\end{abstract}
\maketitle

Recently, the observation of ordered nanostructures such as ripples 
and two-dimensional patterns on ion-sputtered surfaces has attracted 
much attention due to the demonstration of the possibility of fabrication 
of ordered nanoscale structures in a relatively easy and affordable 
way~\cite{books,facsco,frost,ruspo1,girard,naumann,chason}. 
Such experimental results have motivated extensive theoretical 
investigations aiming to understand the mechanism of the morphological 
evolution of ion-sputtered surfaces.
A linear model, proposed by Bradley and Harper (BH)~\cite{bh}, has
been successful in predicting the formation of the ripple
structure. The wavelength, orientation and amplitude of the 
ripples can be predicted in terms of experimental parameters such as 
the incident angle of the ion beam and substrate temperature~\cite{koponen}.
The BH theory, however, fails to explain a number of experimental 
observations such as the saturation of the ripple 
amplitude~\cite{witt,aziz,vajo}, or the appearance of kinetic 
roughening~\cite{eklund,yang}. To remedy such shortcomings, 
the noisy Kuramoto-Sivashinsky (KS) equation~\cite{ks,cuerno} was 
introduced based on the Sigmund theory of sputter erosion~\cite{sigmund}. 
In addition to the linear terms of the BH model, it contains a nonlinear term 
proportional to $(\nabla h)^2$, known as the Kardar-Parisi-Zhang (KPZ) 
term~\cite{kpz}, where $h$ is surface height. Due to the nonlinear term, 
the surface roughness (or the ripple amplitude), which was growing 
exponentially with increasing sputter time in the linear model, 
changes to the type following a power law and eventually saturates to 
a constant value~\cite{park}. Although the KS model seems to be successful 
in offering insights for understanding the nonlinear behavior of sputter-eroded 
surfaces, it has not been convincing yet because the detailed properties of 
the kinetic roughening predicted by the KS equation have not been 
fully tested experimentally.
\par
Kinetic roughening behavior is described by the scaling theory~\cite{fractal}. 
The surface roughness of the sample, $L\times L$ in size, at sputter time $t$ 
is defined as $W\equiv\sqrt{(1/L^2)\sum_\mathbf{r}\big[h(\mathbf{r},t)-
\bar h\big]^2}$, where $\bar h=(1/L^2)\sum_\mathbf{r}h(\mathbf{r},t)$. 
It follows the scaling relation, $W(L,t)\sim L^{\alpha}g(t/L^{\alpha/\beta})$, 
where $g(u)\sim u^{\beta}$ for $u \ll 1$ and $g(u\rightarrow \infty)\sim$ 
constant. The roughness and growth exponents $\alpha$ and $\beta$ 
are related via scaling relations to give the dynamic exponent, 
$z=\alpha/\beta$, which determines the scaling of the saturation time 
with the system size $L$.
Below, we will deal with the inverse dynamic exponent, 
$1/z$, called the coarsening exponent, in analyzing the experimental results.
\par
In this Letter, we study the kinetic roughening of sputter-eroded 
Pd(001) surface via novel experimental techniques combined with 
the stochastic continuum theory. 
Both \textit{in situ} real-time X-ray reflectivity (XRR) and \textit{in situ} 
scanning tunneling microscopy (STM)  show that nanoscale islands 
are formed on surface, evolving with increasing sputter time $t$ following 
the scaling function $W(L,t)$. In particular, we obtain 
$\beta\simeq 0.2$ and $1/z\simeq 0.2$ for ion
beam energy $\epsilon=0.5$ keV, which are not in
agreement with the values predicted by the KS model. 
To resolve this inconsistency, we investigate theoretically 
the erosion process by applying the Sigmund theory to higher order, 
finding that the relevant higher order term is of the form 
$\nabla^2 (\nabla h)^2$, referred to as the conserved KPZ (cKPZ) term. 
Using this new nonlinear term, we can explain the kinetic 
roughening on the sputtered Pd(001) surface successfully. 
\begin{figure}[t]
\centerline{\epsfxsize=7cm \epsfbox{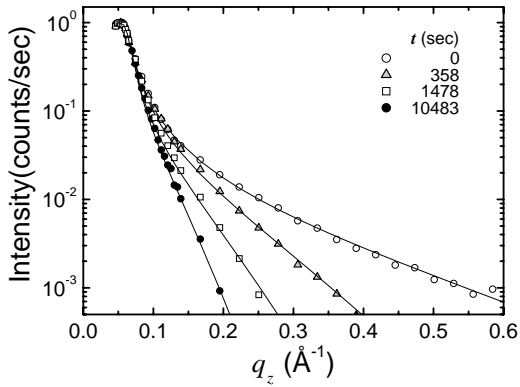}}
\caption{Specular x-ray reflectivity (symbols) as a function of
the momentum transfer, $q_z$, for increasing sputter times 
with $\epsilon=0.5$ keV and $f=2.0\times 10^{13}$ ions/cm$^2$/sec. 
Sputtering is made normal to a sample at room temperature. 
The solid lines are the theoretical ones according to the Parratt formula.}
\label{fig1}
\end{figure}
\begin{figure}[t]
\centerline{\epsfxsize=7cm \epsfbox{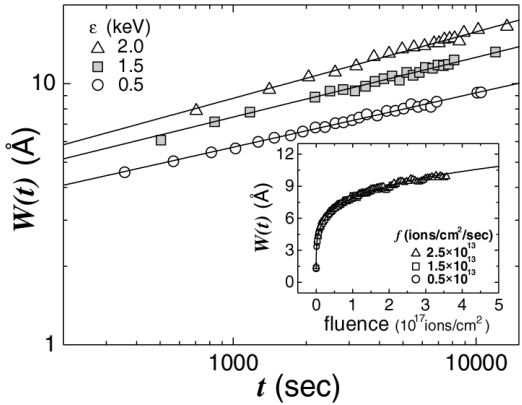}}
\caption{The surface roughness $W$ as a function of sputter time $t$ 
for different ion energies, $\epsilon=0.5(\bigcirc)$, $1.5(\square)$, 
and $2.0(\triangle)$ keV, but with a fixed ionic flux, $f=2.0\times 10^{13}$
ions/cm$^2$/sec. Inset shows that the roughness evolution for 
a given $\epsilon$ follows the same fluence dependence, irrespective of $f$.}
\label{fig2}
\end{figure}

\textit{Experiment.}---A Pd(001) sample was cleaned by several cycles of 
Ar$^+$ ion sputtering around $300$ K and annealing up to $920$ K. 
The clean Pd(001) surface exhibited an average terrace size
of about $3000$ \AA\ and little surface modulation or defects by
contaminants as judged by STM images. The initial surface
roughness of the sample was less than $2$ \AA\ and the miscut from
surface normal direction was $0.3^\circ$, as determined
respectively by X-ray reflectivity (XRR) and X-ray diffraction
spot-profile analysis. Sputtering experiments were performed for
Pd(001) around 300 K with Ar$^+$ ion beam incident normally to the
sample surface. To avoid possible contamination during
sputtering, fresh Ar gas (purity of 99.999\%) was continuously
made to flow through the chamber with Ar partial pressure maintained
around $2.0\times 10 ^{-5}$ Torr.
\par
The morphological evolution of the Pd(001)  during
sputtering was observed \textit{in situ} by both XRR
and STM. The XRR experiment was performed in a custom-designed 
UHV chamber at the
5C2 K-JIST beamline of the Pohang Light Source in Korea. The base
pressure of the chamber was kept below mid-$10^{-10}$ Torr. The
incident X-rays were vertically focused by a mirror to $1$ mm in
size, and a double-bounce Si(111) monochromator was employed to
both focus the beam in the horizontal direction and monochromatize
the incident X-ray to a wavelength of 1.24 \AA\, which
corresponded to the X-ray energy of 10 keV. 
The X-ray reflectivity was measured in real time while
Pd(001)  was sputtered. Incident Ar$^+$ beam energy,
$\epsilon$, ranged from $0.5$ to $2.0$ keV, and ion beam flux, $f$,
measured from the ion current collected at the sample, ranged from
$0.5\times 10^{13}$ to $2.5\times 10^{13}$ ions/cm$^2$/sec.
Sputtering started with the sample at room temperature measured by
a thermocouple directly attached to the back-end of the sample.
After sputtering for more than $10$ hours, the sample temperature rose
only less than $30$ K.
\par
To observe the sputter-induced morphological evolution in
real-space, a commercial STM (Omicron VT-STM) was also used. The
base pressure was kept in the low $10^{-11}$ Torr. STM tips were
electro-chemically etched polycrystalline W-wires that were
annealed and sputtered in a UHV chamber. In all the STM
measurements, the tip-sample voltage was kept as 80 mV, and the
tunneling current was at $1$ nA with the sample at room temperature.
Sputtering was performed with $\epsilon=0.5$ keV and
$f=0.5\times10^{13}$ ions/cm$^2$/sec. During sputtering, the sample
temperature was around 300 K. Sputtering and STM measurement
were alternated, and the sputter time referred to in the present work is
the total sputter time determined by summing all the previous
sputtering periods. 
X-ray reflectivity indicated that there was no significant change 
in the surface roughness for 10 hours after the interruption of
sputtering. Thus, the interruption of sputtering for the STM
measurement was not expected to result in additional relaxation of
the sputtered surface. 
\begin{figure}
\centerline{\epsfxsize=7cm \epsfbox{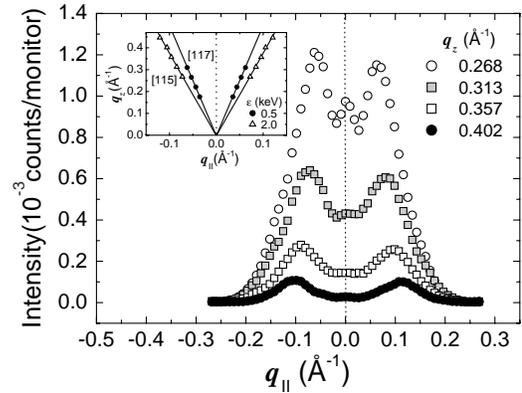}}
\caption{ X-ray diffraction intensity along [110] for different $q_z$.
The positions of satellite peaks shift to the larger $q_\parallel$ for
the larger $q_z$, with a fixed angle to the surface normal, 
indicating that the spots originate from a facet formed by sputtering. 
Inset: Plot of the positions of satellite peaks, $q_\parallel$, as 
a function of $q_z$. From the angle the facet peaks form with respect 
to the surface normal, we identify the facet planes to be (117)  
for $\epsilon=0.5$ keV and (115) for $\epsilon=2.0$ keV.}
\label{fig3}
\end{figure}
\begin{figure}
\centerline{\epsfxsize=7cm \epsfbox{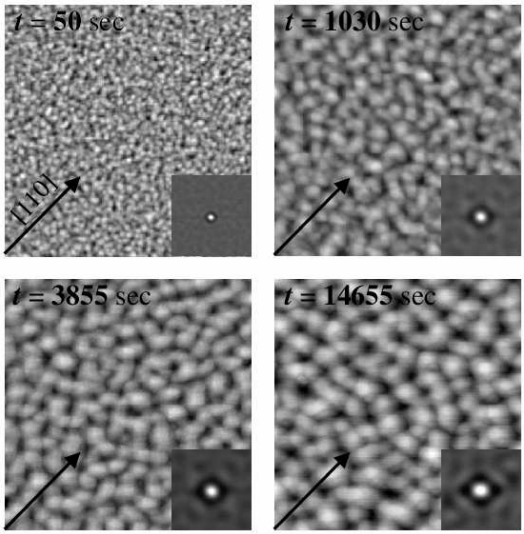}}
\caption{ STM images for various sputtering times which are 
$200\times 200$ nm$^2$ in size. The insets at right bottom of 
each image show the height-height correlation function obtained 
from the corresponding surface profile.}
\label{fig4}
\end{figure}

\textit{Experimental results.}---In Born approximation which holds above critical angle,
the specular reflectivity decreases as $\exp(-W^2 q_z^2)$  
with increasing $q_z$~\cite{Sinha}. Fig.~1 shows that the reflectivity decays more steeply 
as the sputtering proceeds, which indicates that $W(t)$ increases with $t$. 
Quantitatively $W(t)$ is determined by fitting the experimental data 
according to the Parratt formalism~\cite{Sinha,Parratt}. 
Fig.~2 shows $W(t)$ thus obtained, as a function of $t$ for 
several $\epsilon$ but with a fixed $f$. The linear behavior in 
double-logarithmic scale implies power-law behavior, 
$W(t)\sim t^{\beta}$. We obtained that $\beta\simeq 0.20\pm0.02$ 
for $\epsilon=0.5$ keV, $\beta\simeq0.23\pm0.02$ for $\epsilon=1.5$ 
keV, and $\beta\simeq 0.25\pm0.02$ for $\epsilon=2.0$ keV.
\par
The inset of Fig.~2 summarizes the surface roughness, $W(t)$ as 
a function of fluence ($f \times t$) for several different $f$s, 
but with a fixed $\epsilon$. 
We find that the roughness evolves showing the same fluence dependence, 
irrespective of $f$. Such behavior implies that the surface 
diffusion is not induced by thermal activation, but rather by the surface 
sputtering under the present experimental condition~\cite{makeev10}.
\par
We also study the facet formation on the sputtered surface by tracing the 
extra-diffraction peaks or satellites around the diffraction peak 
corresponding to the sample surface, now (001) plane, because  
each facet plane defines diffraction rods normal to it.
In the inset of Fig.~3, the diffraction rods defined by the  satellite
peaks make well-defined angles with respect to that of (001) plane 
as much as the facet plane does to the (001) plane .
From the observed angles, we identify the facet 
formed on sputtering by $0.5$ keV ion beam
to be (117) and that by $2.0$ keV beam to be (115).
\par
The morphological evolution of the sputtered Pd(001) surface were
also investigated by an STM. Fig.~4 shows the STM images of the
sputtered surface at different (total) sputtering time under 
the same experimental condition, $\epsilon=0.5$ keV and 
$f=0.5\times 10^{13}$ ions/cm$^2$. 
The shape of the island edge looks somewhat irregular which
may be due to random shot noise or the sputter-induced diffusion. 
This reaffirms that thermal smoothening is inefficient in the present 
experimental conditions as suggested by the flux-independent scaling behavior.
The insets of Fig.~4 show the height-height correlation function, 
$G(r,t)=\langle (h(r,t)-h(0,0))^2 \rangle$, that reveals lateral order 
developing on the surface as sputtering proceeds. 
We can see that the island size grow with increasing sputter time. 
From the height-height correlation function, we determine the lateral 
correlation length, $\xi$, as the position giving the first-order 
maximum~\cite{correlation}, and plot it as a function of $t$ in Fig.~5. 
It is found that $\xi \sim t^{1/z}$ with the exponent 
$1/z\simeq 0.20\pm0.03$. The growth exponent determined by the STM
measurement is $\beta\simeq 0.20\pm0.03$, 
in agreement with that of the XRR measurement.

\begin{figure}
\centerline{\epsfxsize=7cm \epsfbox{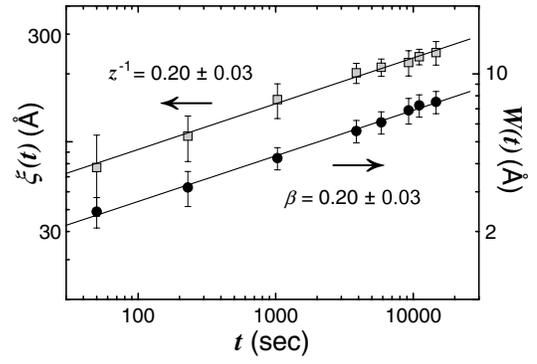}}
\caption{Plot of the lateral correlation length $\xi(\bigcirc)$ 
and the surface roughness $W(\square)$ obtained from the STM images 
as a function of sputter time $t$.}
\label{fig5}
\end{figure}
\begin{figure}
\centerline{\epsfxsize=7cm \epsfbox{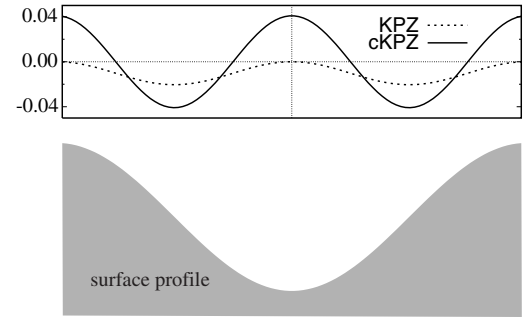}}
\caption{Comparison of the KPZ and the cKPZ terms 
for a prototypical profile of eroded surface.} 
\label{fig6}
\end{figure}

\textit{Theory.}---To elucidate the kinetic roughening, 
we systematically expand the Sigmund 
theory to higher order than that where the KS equation is derived, 
and obtain the dynamic equation, 
\begin{equation}
\frac{1}{c}\frac{\partial h}{\partial t}=
-1-\nu \nabla^2 h -D\nabla^4 h+\lambda_1(\nabla h)^2
+\lambda_2\nabla^2 (\nabla h)^2+\eta(x,y,t)~,
\label{ksckpz}
\end{equation}
where $c$ is the surface recession rate given by the material constant 
times the ionic flux, $\nu$ the effective surface tension generated by 
the erosion process, $D$ the ion-induced effective diffusion 
constant~\cite{makeev10}, $\lambda_1$ and $\lambda_2$ the tilt-dependent 
erosion rate, and $\eta$ an uncorrelated white noise with zero mean, 
mimicking the randomness resulting from the stochastic nature of ion 
arrival at the surface. The coefficients $\nu$, $D$, and $\lambda_1$ 
in Eq.~(\ref{ksckpz}) were derived in terms of experimental parameters 
such as the ion beam energy $\epsilon$, the flux $f$, the incident angle 
$\theta$, the penetration depth $a$, and the cascading sizes in transverse 
and longitudinal directions, $\mu$ and $\sigma$ respectively~\cite{cuerno}. 
The newly derived cKPZ coefficient is, under normal incidence,
given by $\lambda_2=\mu^2/2+(3/8)(\mu/\sigma)^4(\sigma^2-a^2)$. 
Using the {\sc trim} Monte Carlo algorithm~\cite{trim}, all the 
coefficients in Eq.~(\ref{ksckpz}) are determined numerically for 
our experimental parameters, which are tabulated in Table I. 
In Fig.~\ref{fig6}, is shown the estimation of the KPZ and cKPZ terms,  
for a typical profile of the eroded surface.
We find that the cKPZ term should be more relevant to the morphological 
evolution than the KPZ term in a finite system. The contributions from 
other terms allowed by symmetry and of order ${\cal O}(\nabla^4,h^2)$ 
such as $(\nabla^2 h)^2$, are found to be negligible.  
Then, for $\epsilon = 0.5$ keV, the extended KS equation is reduced 
to the cKPZ equation that gives the growth exponent 
$\beta\simeq 0.20$~\cite{ckpz} 
for two dimensions, in agreement with the measured value for
$\epsilon=0.5$ keV. As the ion energy increases,  
the coefficient $D$ increases very rapidly in comparison with 
the magnitudes of the $\lambda_1$ and $\lambda_2$ 
as seen in the Table I. Thus, for the $\epsilon=2.0$ keV, the erosion process 
is mainly governed by the so-called Mullins term, $D\nabla^4 h$~\cite{mullins}. 
In this case, it is known that the growth exponent 
$\beta \simeq 0.25$ for two dimensions, again consistent with the 
experimental value for $\epsilon=2.0$ keV.
The sputter-induced  palladium atoms form adatom islands with well-defined 
facets as observed in Fig.~3.  
Then, $\alpha=1$ and the coarsening exponent, $1/z$, should be equal to 
$\beta$ following the scaling relation $1/z=\beta/\alpha$. 
This prediction is confirmed in the present experiment,  
by the observation of  $1/z=\beta=0.2$   for $\epsilon=0.5$ keV. 
\begin{table}[h]
\begin{tabular}{c|cccc}
\hline
$\epsilon$ & $\nu$ & $D$ & $\lambda_1$ & $\lambda_2$ \\
\hline\hline
~~0.5 keV~~ & ~~2.72~~ & ~~33.35~~ & ~~-0.50~~ & ~~12.2~~ \\
~~1.5 keV~~ & ~~5.12~~ & ~~184.32~~ & ~~-0.54~~ & ~~33.6~~ \\
~~2.0 keV~~ & ~~5.44~~ & ~~266.78~~ & ~~-0.50~~ & ~~49.0~~ \\
\hline
\end{tabular}
\caption{Numerical estimations of the coefficients in Eq.(\ref{ksckpz}) 
under the experimental conditions we performed.}
\end{table}

\textit{Conclusions.}---We have studied the kinetic roughening of 
the sputter-eroded Pd(001) surface both experimentally and 
theoretically. The experimental data suggest that the KS continuum 
model is not relevant to the morphological evolution of the Pd surface. 
Instead, the KS model extended by the inclusion of the cKPZ  term 
we derived from the Sigmund theory could properly describe the kinetic 
roughening behavior of the sputtered Pd(001) surface. 
The model, at the same time, may shed light on recent experimental 
realization of self-organized surface nanostructures induced by ion sputtering.
For example, the parameter space for nanoscale island formation predicted by 
the KS model, which seems too narrow to be realistic as compared with 
experimental observations~\cite{qdots}, can be enhanced. 

\begin{acknowledgements}
This work is supported by the Korean Research Foundation
(KRF-2002-015-CP0087) through SMU and KOSEF Grant
No.2002-2-11200-002-3 in the BRP program through SNU.
\end{acknowledgements}

\end{document}